\documentclass[twocolumn,reprint,floatfix,nofootinbib]{revtex4-1}
\usepackage{amsmath,amssymb}
\usepackage[dvipdfmx]{graphicx}
\usepackage{bm}
\usepackage{braket}
\usepackage{physics}
\usepackage{color}
\usepackage{natbib}
\usepackage[pdfencoding=auto]{hyperref}
\usepackage{comment}

\begin{document}

\title{Spin texture and spin current in excitonic phases of the two-band Hubbard model} 
\author{
Hisao Nishida,$^{1}$
Shohei Miyakoshi,$^{2}$
Tatsuya Kaneko,$^{3}$
Koudai Sugimoto,$^{4}$
and Yukinori Ohta$^{1}$
}
\affiliation{
$^{1}$Department of Physics, Chiba University, Chiba 263-8522, Japan\\
$^{2}$Computational Quantum Matter Research Team, RIKEN Center for Emergent Matter Science (CEMS), Wako, Saitama 351-0198, Japan\\
$^{3}$Computational Condensed Matter Physics Laboratory, RIKEN Cluster for Pioneering Research (CPR), Wako, Saitama 351-0198, Japan\\
$^{4}$Center for Frontier Science, Chiba University, Chiba 263-8522, Japan
}

\date{\today}

\begin{abstract}
Using the mean-field approximation, we study the $k$-space spin textures and local spin currents 
emerged in the spin-triplet excitonic insulator states of the two-band Hubbard model defined on 
the square and triangular lattices.  We assume a noninteracting band structure with a direct band gap 
and introduce $s$-, $p$-, $d$-, and $f$-type cross-hopping integrals, i.e., the hopping of electrons 
between different orbitals on adjacent sites with four different symmetries.  
First, we calculate the ground-state phase diagrams in the parameter space of the band filling and 
interaction strengths, whereby we present the filling dependence of the amplitude and phase of the 
excitonic order parameters.  Then, we demonstrate that the spin textures (or asymmetric band structures) 
are emerged in the Fermi surfaces by the excitonic symmetry breaking when particular phases 
of the order parameter are stabilized.  Moreover, in case of the $p$-type cross-hopping integrals, 
we find that the local spin current can be induced spontaneously in the system, which does not contradict 
the Bloch theorem for the absence of the global spin current.  The proofs of the absence of the global 
spin current and the possible presence of the local spin currents are given on the basis of the Bloch 
theorem and symmetry arguments.  
\end{abstract}

\maketitle

\section{Introduction}

The excitonic phase, which is sometimes referred to as the excitonic insulator phase, is the state where 
valence-band holes and conduction-band electrons in small band-gap semiconductors or small band-overlap 
semimetals form pairs (or excitons) due to weakly screened Coulomb interactions, and a macroscopic number 
of the pairs condense into a quantum state acquiring the phase coherence.  Although the excitonic phase was 
predicted to occur more than half a century ago as a spontaneous hybridization between the valence and conduction 
bands and has attracted much attention because a theoretical framework similar to that of BCS superconductors 
can be applied \cite{KK65, De65, JRK67, Ko67, HR68_1, HR68_2, PhysRevB.74.165107}, the lack of candidate materials 
delayed our understandings of this phase until recently.  However, the progress in this research field has been made 
rapidly in recent years owing to the discovery of some candidate materials.  The spin-singlet excitonic phase has 
been suggested to emerge in some transition-metal chalcogenides such as 1$T$-TiSe$_2$ 
\cite{CMCetal07,MCCetal09,ZFBetal13,MMHetal15,WSY15,PhysRevB.97.155131} and Ta$_2$NiSe$_5$ 
\cite{WSTetal09, KTKetal13, SWKetal14, SKO16, YDO16, SO16, LKLetal17, SNKO18}, 
and the spin-triplet excitonic phase has also been suggested to emerge in some cobalt oxide materials 
located in the crossover regime between the high-spin and low-spin states 
\cite{KA14-2, INMetal16, NWNetal16, SK16, TMNetal16, YSO17, PhysRevB.95.115131, PhysRevB.96.245102, 
PhysRevB.98.205105, 2018arXiv181103779I}.  Since these materials are transition-metal compounds, 
the relevant properties should be considered within the framework of the physics of strong electron 
correlations using the Hubbard-like lattice models 
\cite{PhysRevB.78.193103, PhysRevB.81.205117, SEO11, PhysRevB.85.121102, EKOetal14, hamada2017excitonic, 
KO14, KZFetal15, Ku14, Ku15, KO16}.  

In a series of such studies, Kune\v{s} and Geffroy \cite{KG16} discussed the effects of cross-hopping integrals 
on the excitonic states in the two-band Hubbard model, where the cross hopping is defined as 
the hopping of electrons between different orbitals on the adjacent sites.  The hopping integral between 
different orbitals on the same site vanishes exactly because of the orthogonality of the orbitals, but the 
cross-hopping integrals between the adjacent sites can have a finite value \cite{KA14-1}.  Since the 
hybridization between the orthogonal orbitals occurs spontaneously due to interorbital Coulomb interactions 
in the excitonic phase, one may naturally expect that the hybridization caused by the cross-hopping integrals 
should affect the excitonic phase significantly.  Kune\v{s} and Geffroy, in particular, showed that the $k$-space 
spin texture, similar to the one derived from the Rashba-Dresselhaus spin-orbit coupling, can appear in the 
spin-triplet excitonic phases even in centrosymmetric lattices without any intrinsic spin-orbit coupling.  

Kune\v{s} and Geffroy \cite{KG16} also argued that the spontaneous spin currents can appear if the order 
parameters of the spin-triplet excitonic phase are imaginary.  Using different models with certain cross-hopping 
integrals, Volkov {\it et al.} \cite{VK78, VGKetal81} discussed the relationship between the excitonic phase 
and imaginary order parameters and showed that the spin current of the orbital off-diagonal components 
can remain finite, but the total spin current including both the orbital diagonal and off-diagonal components 
vanishes exactly.  Thus, they concluded that the global spin currents can never appear spontaneously in 
the equilibrium excitonic phase.  Geffroy {\it et al.}~\cite{PhysRevB.97.155114} also pointed out the absence 
of the global spin current.  This result is consistent with the Bloch theorem \cite{ohashi1996bloch, tada2016two} 
that claims that the global spin current does not appear spontaneously in the ground state.  The existence of 
the spontaneous global spin current is thus unlikely to occur in the excitonic phases of strongly 
correlated electron systems.  

In this paper, motivated by the above developments in the field, we study the excitonic phases of the 
two-band Hubbard models with cross-hopping integrals within the mean-field approximation.  
We assume the square and triangular lattices in two-dimension and examine the cross-hopping integrals 
of four types, i.e., $s$-, $p$-, $d$-, and $f$-types.  We thus calculate the ground-state phase diagram of the 
models and clarify the behaviors of the excitonic order parameters, Fermi surfaces, band dispersions, 
and spin currents.  In particular, we discuss the relationship between the global spin currents and 
excitonic phase with imaginary order parameters.  We thereby find that the spin textures are emerged in the 
Fermi surfaces by the excitonic symmetry breaking when particular phases of the order parameter are 
stabilized and that the local spin current can be induced in the system with the $p$-type cross-hopping 
integral.  The proofs of the absence of the global spin current and the possible presence of the local spin 
currents are also given on the basis of the Bloch theorem and symmetry arguments.  We thus present a 
comprehensive understanding of the spin textures and spin currents in the spin-triplet excitonic phases 
of the two-band Hubbard model.  

The rest of this paper is organized as follows.  
In Sec.~II, we introduce the two-band Hubbard model with the cross hopping integrals and derive the 
self-consistent equations for obtaining the ground state of the model in the mean-field approximation.  
In Sec.~III, we present the calculated results for the phase diagram of the system, $k$-space spin texture, 
features of the order parameters, and the local and global spin currents of the system.  
We summarize our results in Sec.~IV.  
Appendices are provided to show the proofs of the absence of the global spin currents and the possible 
presence of the local spin currents in the excitonic phases of the model. 

%----------------------------------------------------------------------------------------%fig
\begin{figure}[thb]
\begin{center}
\includegraphics[keepaspectratio=true,width=0.8\columnwidth]{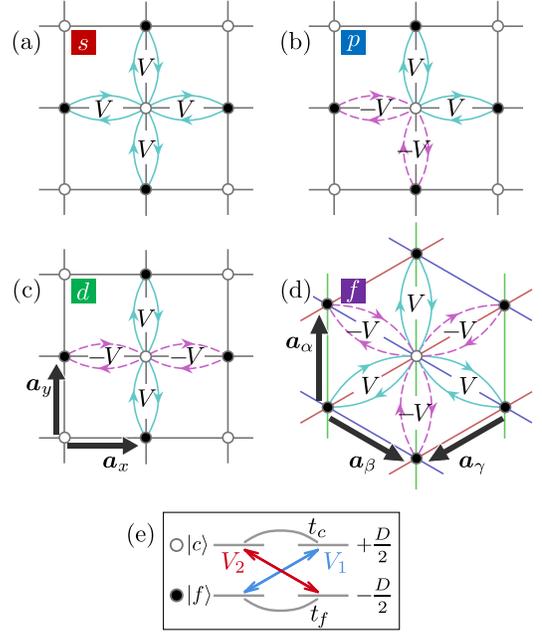} 
\end{center}
\caption{Schematic representations of (a) the $s$-, (b) $p$-, (c) $d$-, and (d) $f$-type cross-hopping 
integrals defined on either the square or triangular lattice.  The blue bonds indicate $V_{1, \tau} = V_{2, \tau} = V$ 
and the purple bonds indicate $V_{1, \tau} = -V_{2, \tau}=V$.  In (e), the direct hoppings ($t_c$ and $t_f$) 
and cross hoppings ($V_1$ and $V_2$) are illustrated.  
The primitive translation vectors are 
$\bm{a}_x=(1,0)$ and 
$\bm{a}_y=(0,1)$ in (c), and 
$\bm{a}_\alpha=(0,1)$, 
$\bm{a}_\beta=(\sqrt{3}/2,-1/2)$, and 
$\bm{a}_\gamma=(-\sqrt{3}/2,-1/2)$ in (d).  
}\label{fig:fig1}
\end{figure}
%----------------------------------------------------------------------------------------%fig

\section{Model and method}

%%%%%%%%%%%%%%%%%%%%%%%%%%%%%%%%
\subsection{Model}
%%%%%%%%%%%%%%%%%%%%%%%%%%%%%%%%

We consider the two-band Hubbard model defined on the two-dimensional lattices.  
The Hamiltonian is written as 
\begin{align}
&\mathcal{\hat{H}}
= \mathcal{\hat{H}}_{\mathrm{t}} + \mathcal{\hat{H}}_{\mathrm{int}} ,
\\
&\mathcal{\hat{H}}_\mathrm{t}
= \sum_{j,\tau,\sigma} \left( t_c \hat{c}^{\dagger}_{j+\tau, \sigma} \hat{c}_{j,\sigma}
		+ t_{f} \hat{f}_{j+\tau,\sigma}^{\dagger} \hat{f}_{j, \sigma} + \mathrm{H.c.} \right)
\notag \\
&~~~~~~+\sum_{j,\tau,\sigma}
\left( V_{1,\tau}\hat{c}^\dagger_{j + \tau, \sigma} \hat{f}_{j,\sigma}
+ V_{2,\tau}f^\dagger_{j + \tau, \sigma} \hat{c}_{j,\sigma} + \mathrm{H.c.} \right) 
\notag \\
&~~~~~~+ \frac{D}{2}\sum_{j,\sigma} \left( \hat{n}^{c}_{j,\sigma}-\hat{n}^{f}_{j,\sigma} \right)
-\mu\sum_{j,\sigma} \left( \hat{n}^{c}_{j, \sigma} + \hat{n}^{f}_{j, \sigma} \right) ,
\label{eq:Ht}
\\
&\mathcal{\hat{H}}_\mathrm{int}=
\sum_{j} \left( U_{c}\hat{n}^{c}_{j,\uparrow}\hat{n}^{c}_{j,\downarrow}+U_{f}\hat{n}^{f}_{j,\uparrow}\hat{n}^{f}_{j,\downarrow}\right)
\notag \\
&~~~~~~+U'\sum_{j,\sigma,\sigma'}\hat{n}^{c}_{j,\sigma}\hat{n}^{f}_{j,\sigma'} , 
\label{eq:Hint}
\end{align}
where $\hat{c}_{j,\sigma}^{\dagger}$ ($\hat{f}_{j,\sigma}^{\dag}$) and $\hat{c}_{j,\sigma}^{}$ ($\hat{f}_{j,\sigma}^{}$) 
are the creation and annihilation operators of an electron on the conduction ($c$) orbital 
[valence ($f$) orbital] at site $j$ with spin $\sigma$.  We define the number operators 
$\hat{n}^{c}_{j,\sigma} = \hat{c}_{j,\sigma}^{\dagger} \hat{c}_{j,\sigma}$ and 
$\hat{n}^{f}_{j,\sigma} = \hat{f}_{j,\sigma}^{\dagger} \hat{f}_{j,\sigma}$.  
In $\mathcal{\hat{H}}_{\mathrm{t}}$, 
$D$ is the on-site energy splitting, 
$\mu$ is the chemical potential, 
% $j + \tau$ is the nearest-neighbor site of $j$, 
$t_c$ and $t_f$ are the hopping integrals between the same orbitals on the nearest-neighbor sites, 
and $V_{1, \tau}$ and $V_{2, \tau}$ are the hopping integrals between the different orbitals on the nearest-neighbor sites.  
$V_{1, \tau}$ and $V_{2, \tau}$ are referred to as the cross-hopping integrals.  
Note that $j+\tau$ indicates the nearest-neighbor site of $j$, where $j$ runs over all sites in the system 
and $\tau$ denotes the primitive translation vector $\bm{a}_\tau$ illustrated in Fig.~\ref{fig:fig1}(c) for the 
square lattice and Fig.~\ref{fig:fig1}(d) for the triangular lattice. 
In $\mathcal{\hat{H}}_{\rm int}$, $U_c$ and $U_f$ are the intraorbital Coulombic repulsive interactions, 
and $U'$ is the interorbital Coulombic repulsive interaction.  
This model is illustrated in Fig.~\ref{fig:fig1}, where $\bm{a}_{\tau}$ is the vector from site $j$ 
to site $j+\tau$ (or the primitive translation vector).  The Fourier transformation of Eq.~(\ref{eq:Ht}) reads 
\begin{align}
{\cal \hat{H}}_\mathrm{t}=\sum_{{\bm k},\sigma}
\begin{pmatrix}{\hat{c}}_{{\bm k},\sigma}^{\dagger} & {\hat{f}}_{{\bm k},\sigma}^{\dagger}\end{pmatrix}
\renewcommand\arraystretch{1.}
\left(\begin{array} {cc}
{\varepsilon}_{c}(\bm{k})  &  {\gamma}(\bm{k})\\
{\gamma}^{*}(\bm{k}) &{\varepsilon}_{f}(\bm{k})
\end{array} \right)
\begin{pmatrix}
{\hat{c}}_{{\bm k},\sigma}^{\ }\\ {\hat{f}}_{{\bm k},\sigma}^{\ }
\end{pmatrix} , 
\end{align}
where the matrix elements are
\begin{align}
{\varepsilon}_{c}(\bm{k})&=2t_c\sum_\tau \cos{k_\tau}+\frac{D}{2}-\mu , \\
{\varepsilon}_{f}(\bm{k})&=2t_f\sum_\tau \cos{k_\tau}-\frac{D}{2}-\mu , \\
{\gamma}(\bm{k})&=2\sum_\tau (V_\tau \cos{k_\tau}+iV^\prime_\tau \sin{k_\tau}) , 
\label{eq:gamma}
\end{align}
with $V_{\tau}=(V_{1,\tau}+V_{2,\tau})/2$, $V^\prime_\tau=(V_{1,\tau}-V_{2,\tau})/2$, and 
$k_\tau={\bm k}\cdot\bm{a}_\tau$.  We assume the hopping integrals as $-t_c = t_f = t = 1$ 
(direct gap) and set $D = 6$ and $U_c = U_f = U$ throughout the paper.  

We consider four types of the cross-hopping integrals, i.e., $s$-, $p$-, $d$-, and $f$-types, 
where $s$-, $p$-, and $d$-types are for the square lattice and $f$-type is for the triangular lattice.  
The signs of $V_{1,\tau}$ and $V_{2,\tau}$ for each type are shown in Fig.~\ref{fig:fig1}.  
We set $V_{1,\tau}=V_{2,\tau}=V_{1,-\tau} = V_{2,-\tau}$ for the $s$- and $d$-types, 
and $V_{1,\tau}= - V_{2,\tau} = -V_{1,-\tau}= V_{2,-\tau}$ for the $p$- and $f$-types, 
where $-\tau$ denotes the primitive translation vector in the opposite direction, $-\bm{a}_\tau$.  
Thus, we rewrite Eq.~(\ref{eq:gamma}) as
\begin{equation}
 \gamma(\bm{k}) = 2 \left( V_{x} \cos{k_x} + V_{y} \cos{k_y} \right)
\end{equation}
for $s$-type,
\begin{equation}
 \gamma(\bm{k}) = 2i \left( V^\prime_{x} \sin{k_x} + V^\prime_{y} \sin{k_y} \right)
\end{equation}
for $p$-type,
\begin{equation}
 \gamma(\bm{k}) = 2 \left( V_{x} \cos{k_x} - V_{y} \cos{k_y} \right)
\end{equation}
for $d$-type, and
\begin{equation}
 \gamma(\bm{k}) = 2i \displaystyle \sum_{\tau=\alpha,\beta,\gamma}V^\prime_{\tau}\sin{k_\tau}
\end{equation}
with $k_{\gamma}=-k_{\alpha}-k_{\beta}$ for $f$-type.  
Hereafter, we assume $V_{\tau} = V'_{\tau} = V = 0.1t$.  
Note that the space inversion of the $s$- and $d$-type cross-hopping integrals has even parity, 
while that of the $p$- and $f$-type ones has odd parity.  
Also, when there are no cross-hopping integrals, the ground state of our two-band model 
at half filling is a band insulator for $U', D \gg U$, while it is a Mott insulator for $U^\prime, D\ll U$, 
and the excitonic insulator state appears in the intermediate region \cite{ZTB11,KSO12,KA14-1}.  
%----------------------------------------------------------------------------------------%fig
%add 2019/1/11 下に移動
%----------------------------------------------------------------------------------------%fig

%%%%%%%%%%%%%%%%%%%%%%%%%%%%%%%%
\subsection{Mean-field theory}
%%%%%%%%%%%%%%%%%%%%%%%%%%%%%%%%

We use the mean-field theory to obtain the ground state of the model.  The excitonic order 
parameter is given by 
\begin{align}
\Phi_{\bm{q}}=&
\frac{1}{L^2}\sum_{j,\sigma,\sigma'}
e^{-i\bm{q}\cdot\bm{r}_{j}}
\expval{\hat{c}^{\dagger}_{j,\sigma}
T_{\sigma,\sigma'}(l)\hat{f}_{j,\sigma'}}
\nonumber\\=&
\frac{1}{L^2}\sum_{\bm{k},\sigma,\sigma'}
\expval{\hat{c}^{\dagger}_{\bm{k}+\bm{q},\sigma}
T_{\sigma,\sigma'}(l)
\hat{f}_{\bm{k},\sigma'}}, 
\end{align}
where $T(l)=l_{0}I+\bm{l}\bm{\cdot}\bm{\sigma}$ with $l=(l_{0},\bm{l})$ and $\bm{l}=(l_1,l_2,l_3)$, 
satisfying $l_0^2+\bm{l\cdot l}=1$ for real numbers $l_r$ $(r=0,1,2,3)$.  
$L^2$ is the number of lattice sites in the system.  
In this paper, we assume the spin-triplet excitonic order of the spin direction along $z$-axis: 
i.e., $T(l)=\sigma_z$.  Note that the energy of the spin-singlet excitonic order ($l_0 \neq 0$ and $\bm{l} = 0$) 
and that of the spin-triplet excitonic order ($l_0 = 0$ and $\bm{l} \neq 0$) are the same in the present model.  
However, we implicitly assume the presence of the exchange interactions like Hund's rule coupling, which 
stabilizes the spin-triplet excitonic order \cite{KO14}.  We do not consider the spin-singlet excitonic order, 
which may be stabilized in the presence of strong electron-phonon couplings \cite{KZFetal15}.  

If we restrict ourselves to the case $\bm{q}=0$ (direct gap), the excitonic ordering changes 
the matrix $\gamma(\bm{k})$ as 
\begin{align}
\gamma(\bm{k})\rightarrow \gamma^\prime_\sigma(\bm{k})=\gamma(\bm{k})- \frac{U^\prime}{2}\sigma\Phi^{*}_{0}. 
\label{eq:eq-tx-del}
\end{align}
The symmetry of the excitonic order depends on the phases of the hybridization term 
$\gamma(\bm{k})$ and order parameter $\Phi_0$.  
The mean-field Hamiltonian of the two-band Hubbard model may then read 
\begin{align}
{\cal \hat{H}}_{\rm MF} \!=\! \sum_{\bm{k},\sigma}
\begin{pmatrix}{\hat{c}}_{{\bm k},\sigma}^{\dagger} \! & \! {\hat{f}}_{{\bm k},\sigma}^{\dagger}\end{pmatrix}
\left(\begin{array} {cc}
{\varepsilon}^\prime_{c}(\bm{k})  &  {\gamma_\sigma^\prime}(\bm{k})\\
{\gamma_\sigma^\prime}^{*}(\bm{k}) &{\varepsilon}^\prime_{f}(\bm{k})  
\end{array} \right)
\begin{pmatrix}
{\hat{c}}_{{\bm k},\sigma}^{\ }\\ {\hat{f}}_{{\bm k},\sigma}^{\ }
\end{pmatrix}
\!+2L^2\varepsilon_0,
\end{align}
with 
\begin{align}
&{\varepsilon_c^\prime}(\bm{k})=2g_c({\bm k})+\frac{D}{2}-\frac{n}{4}(U-2U^\prime)-\mu_0 ,
\\
&{\varepsilon_f^\prime}(\bm{k})=2g_f({\bm k})-\frac{D}{2}+\frac{n}{4}(U-2U^\prime)-\mu_0 ,
\\
&{\gamma_\sigma^\prime}(\bm{k})=2 h(\bm{k})-\frac{U^\prime}{2}\sigma |\Phi_0^\mathrm{t}|e^{-i\phi} , 
\label{eq:gamma^p}
\\
&\varepsilon_0=-\left(\frac{N}{4}\right)^2(U+2U^\prime)-
\left(\frac{n}{4}\right)^2(U-2U^\prime)+\frac{U^\prime}{4}|\Phi^t_0|^2, 
\end{align}
where we define $g_{c(f)}(\bm{k})=t_{c(f)}\sum_{\tau}\cos k_{\tau}$, 
$h({\bm k})=\sum_{\tau}(V_{\tau}\cos k_\tau+iV^\prime_\tau\sin k_\tau)$, and 
$\mu_0=\mu-\frac{N}{4}(U+2U^\prime)$.  
The number of electrons per unit cell is given by 
$N=\frac{1}{L^2}\sum_{\bm{k},\sigma}\big(\langle{\hat{n}}_{\bm{k},\sigma}^{f}\rangle+\langle{\hat{n}}_{\bm{k},\sigma}^{c}\rangle\big)$, 
where $N=2$ is for the half-filled band, and the difference between the numbers of $c$ and $f$ electrons is given by 
$n=\frac{1}{L^2}\sum_{\bm{k},\sigma}\big(\langle{\hat{n}}_{\bm{k},\sigma}^{f}\rangle-\langle{\hat{n}}_{\bm{k},\sigma}^{c}\rangle\big)$.  

We define the $\bm{q} = 0$ spin-triplet excitonic order parameter as 
\begin{equation}
\Phi_0^\mathrm{t}=|\Phi_0^\mathrm{t}|e^{i\phi} 
=\frac{1}{L^2}\sum_{\bm{k},\sigma}\sigma \langle {\hat{c}}_{\bm{k},\sigma}^{\dagger}{\hat{f}}_{\bm{k},\sigma}^{\  }\rangle, 
\end{equation}
where $\phi$ is the phase of the complex order parameter and $\sigma = \pm 1$.  
The mean-field Hamiltonian can be diagonalized by the Bogoliubov transformation 
\begin{align}
\begin{pmatrix}
{\hat{\alpha}}_{{\bm k},\sigma,+}\\
{\hat{\alpha}}_{{\bm k},\sigma,-}
\end{pmatrix}
=\left(\begin{array} {cc}
 u_{\bm{k},\sigma}   &e^{i\theta_{\bm{k},\sigma}}v_{\bm{k},\sigma}\\
 e^{-i\theta_{\bm{k},\sigma}}v_{\bm{k},\sigma} &-u_{\bm{k},\sigma}
\end{array} \right)
\begin{pmatrix}
{\hat{c}}_{{\bm k},\sigma}^{\ }\\ {\hat{f}}_{{\bm k},\sigma}^{\ }
\end{pmatrix} , 
\label{eq:bogoliubov}
\end{align}
where we take $u_{\bm{k},\sigma}$ and $v_{\bm{k},\sigma}$ to be real, 
and $v_{\bm{k},\sigma}$ is multiplied by the phase factor $e^{i \theta_{\bm{k},\sigma}}$.  
This assumption does not lose generality because the relative phase of 
$u_{\bm{k},\sigma}$ and $v_{\bm{k},\sigma}$ is fixed in the Bogoliubov transformation.  
Since the Bogoliubov transformation is unitary, the identity 
$\left| u_{\bm{k},\sigma} \right|^2 + \left| v_{\bm{k}, \sigma} \right|^2 = 1$ 
is satisfied.  Thus, we obtain the diagonalized mean-field Hamiltonian as 
\begin{align}
{\cal \hat{H}}_{\rm MF} \! = \! \sum_{{\bm k},\sigma}\left(
{E}^{+}_{{\bm k},\sigma}{\hat{\alpha}}_{{\bm k},\sigma,+}^{\dagger}{\hat{\alpha}}_{{\bm k},\sigma,+}^{\ }
\!+\! {E}^{-}_{{\bm k},\sigma}\hat{\alpha}_{{\bm k},\sigma,-}^{\dagger}{\hat{\alpha}}_{{\bm k},\sigma,-}^{\ } \!
\right) \!+\! 2L^2\varepsilon_0 , 
\end{align}
with the quasiparticle band dispersions 
\begin{align}
{E}_{\bm{k},\sigma}^{\pm}&=\eta_{\bm k} \pm \sqrt{ \xi_{\bm k}^2+|\gamma_\sigma^\prime(\bm{k})|^2  } ,
\label{eq:Epm}\\
\eta_{\bm k}&=\frac{1}{2}\big( {\varepsilon_c^\prime}(\bm{k})+{\varepsilon_f^\prime}(\bm{k}) \big) , \\
\xi_{\bm k}&=\frac{1}{2}\big( {\varepsilon_c^\prime}(\bm{k})-{\varepsilon_f^\prime}(\bm{k}) \big). 
\end{align}
The transformation coefficients and complex phase factor are given by 
\vspace*{-0.1cm} 
\begin{align}
&u_{\bm{k},\sigma}^2=\frac{1}{2}\left(1+\frac{\xi_{\bm k}}{\sqrt{ \xi_{\bm k}^2+|\gamma_\sigma^\prime(\bm{k})|^2}}\right) ,
\label{eq: bogoliubov-uk} \\
&v_{\bm{k},\sigma}^2=\frac{1}{2}\left(1-\frac{\xi_{\bm k}}{\sqrt{ \xi_{\bm k}^2+|\gamma_\sigma^\prime(\bm{k})|^2}}\right) ,
\label{eq: bogoliubov-vk} \\
& e^{i\theta_{\bm{k},\sigma}}= \frac{\gamma_\sigma^\prime(\bm{k})}{|\gamma_\sigma^\prime(\bm{k})|} . 
\label{eq: bogoliubov-phase}
\end{align}
Thus, the self-consistent equations are given by 
\begin{align}
&N = \frac{1}{L^2}\sum_{{\bm k},\sigma} \left[ f(E_{\bm{k},\sigma}^{+})+f(E_{\bm{k},\sigma}^{-}) \right] ,
\\
&n = \frac{1}{L^2}\sum_{{\bm k},\sigma}
\left( u_{\bm{k},\sigma}^2-v_{\bm{k},\sigma}^2 \right)  \left[ f(E_{\bm{k},\sigma}^{+})-f(E_{\bm{k},\sigma}^{-}) \right] ,
\\
&\Phi_0^\mathrm{t} = \frac{1}{L^2}\sum_{{\bm k},\sigma} \sigma
u_{\bm{k},\sigma}v_{\bm{k},\sigma} e^{-i\theta_{\bm{k},\sigma} } \left[ f(E_{\bm{k},\sigma}^{+})-f(E_{\bm{k},\sigma}^{-}) \right] ,  
\end{align}
where we define the Fermi distribution function $f(E_{\bm{k}, \sigma}^{\pm}) 
=\langle {\hat{\alpha}}_{\bm{k},\sigma,\pm}^{\dagger} {\hat{\alpha}}_{\bm{k}, \sigma, \pm} \rangle
=1/(1+e^{\beta E_{\bm{k}, \sigma}^{\pm}})$ using the reciprocal temperature $\beta$.  
We carry out the following calculations at zero temperature.  

%\vfill

%----------------------------------------------------------------------------------------%fig
\begin{figure*}[thb]
\begin{center}
\includegraphics[keepaspectratio=true,width=1.9\columnwidth]{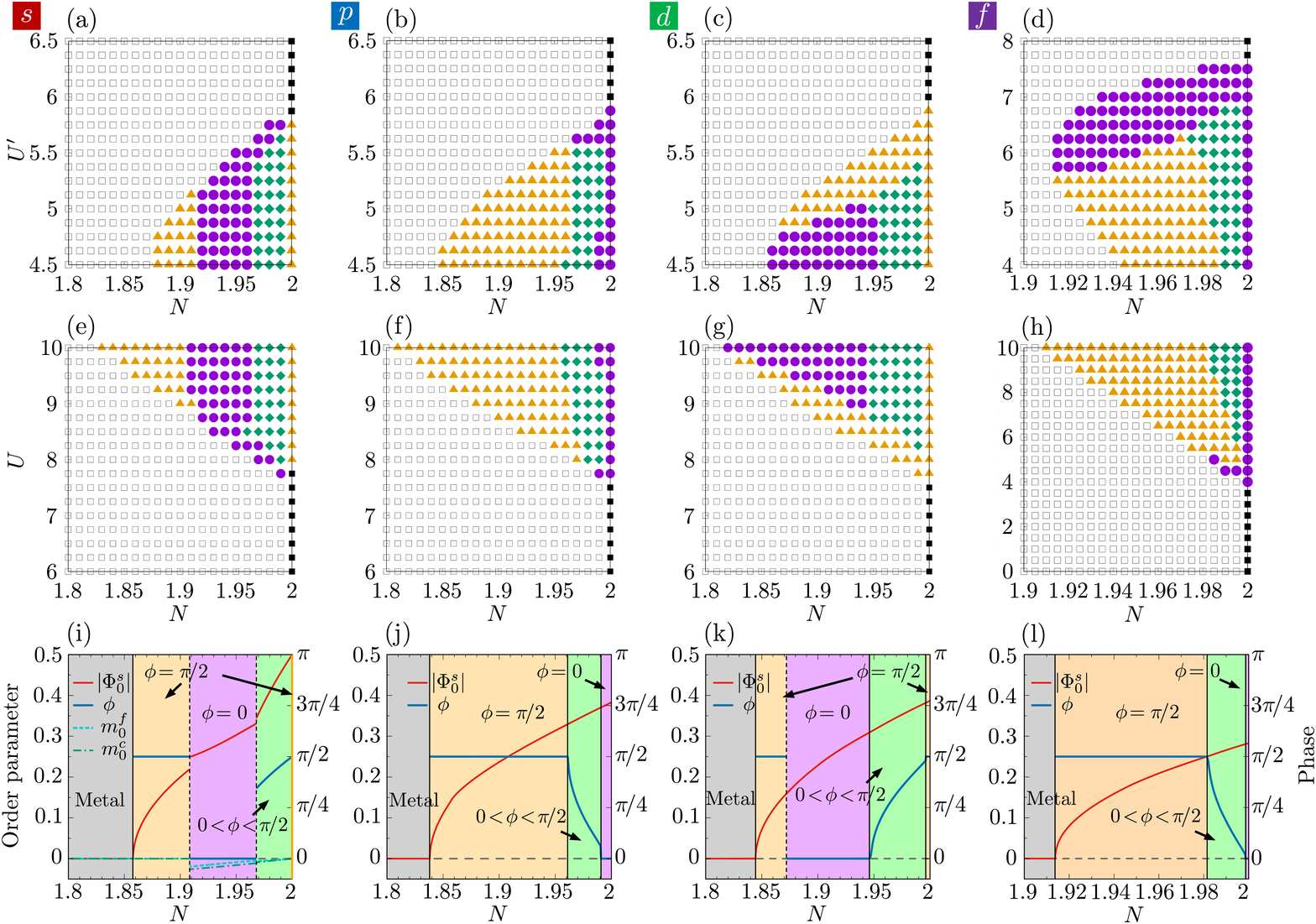}
\end{center}
\vspace*{-0.2cm} %add 2019/1/11
\caption{Calculated phase diagrams of our model in the parameter space of $(N,U')$ at $U=9$ (upper panels) 
and $(N,U)$ at $U'=5$ (middle panels), where $N$ is the number of electrons per site.  The cross-hopping 
integrals of (a,e) $s$-, (b,f) $p$-, (c,g) $d$-, and (d,h) $f$-type are assumed.  Circles, diamonds, and triangles 
in the phase diagrams represent the excitonic phases with the phase $\phi =0$, $0< \phi < \pi/2$, and 
$\phi =\pi/2$, respectively, and squares represent the normal phase.  
In the lower panels (i)-(l), we show the calculated amplitude $|\Phi_0^\mathrm{t}|$ (red) and phase $\phi$ (blue) of the 
excitonic order parameter at $U = 9.5$ and $U'=5$ as a function of $N$, where we assume the cross-hopping 
parameters of (i) $s$-, (j) $p$-, (k) $d$-, and (l) $f$-type.  The solid and dotted lines at the phase boundaries 
represent the second- and first-order phase transitions, respectively.  
}\label{fig:fig2}
\end{figure*}
%----------------------------------------------------------------------------------------%fig

%----------------------------------------------------------------------------------------%fig
\begin{figure}[thb]
\begin{center}
\includegraphics[keepaspectratio=true,width=0.9\columnwidth]{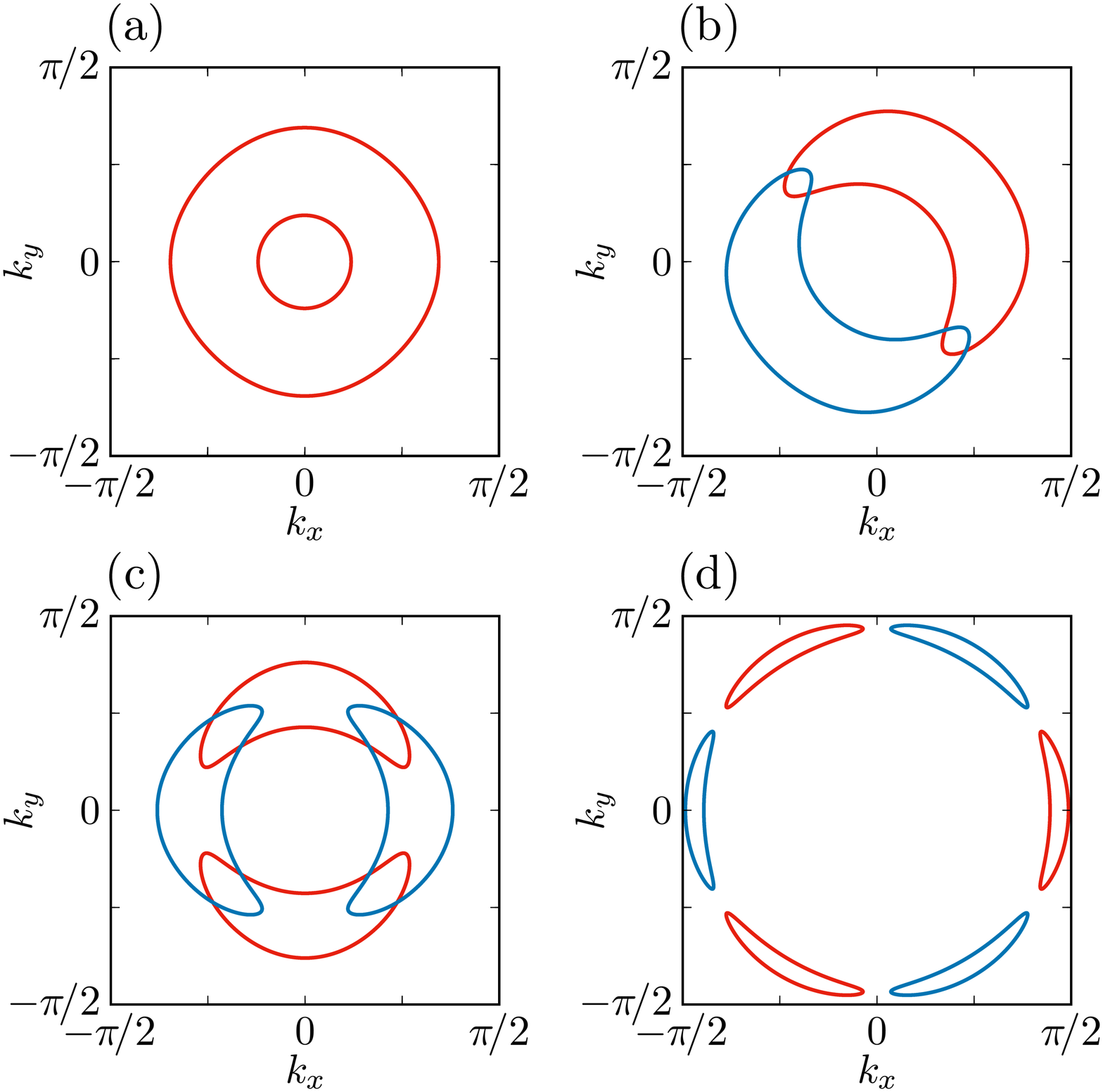}
\end{center}
\vspace*{-0.4cm} %add 2019/1/11
\caption{Calculated $k$-space spin textures for the (a) $s$-, (b) $p$-, (c) $d$-, and (f) $f$-type 
cross-hopping integrals, where the up-spin (red) and down-spin (blue) Fermi surfaces are drawn.  
We assume $N=1.94$ in (a), (b), and (c), and $N=1.98$ in (d).  We find $\phi=0$ in (a) and (c) 
and $\phi=\pi/2$ in (b) and (d).  We set $D=6$ , $U=9.5$, and $U^\prime=5$.  
The areas enclosed by the dotted lines in Figs.~\ref{fig:fig4}(g) and \ref{fig:fig4}(h) 
(see below) are shown.  
}\label{fig:fig3}
\end{figure}
%----------------------------------------------------------------------------------------%fig

\section{Results and discussion}

%%%%%%%%%%%%%%%%%%%%%%%%%%%%%%%%
\subsection{Order parameter and $k$-space spin textures}
%%%%%%%%%%%%%%%%%%%%%%%%%%%%%%%%

\begin{figure}[thb]
\begin{center}
\includegraphics[keepaspectratio=true,width=1.0\columnwidth]{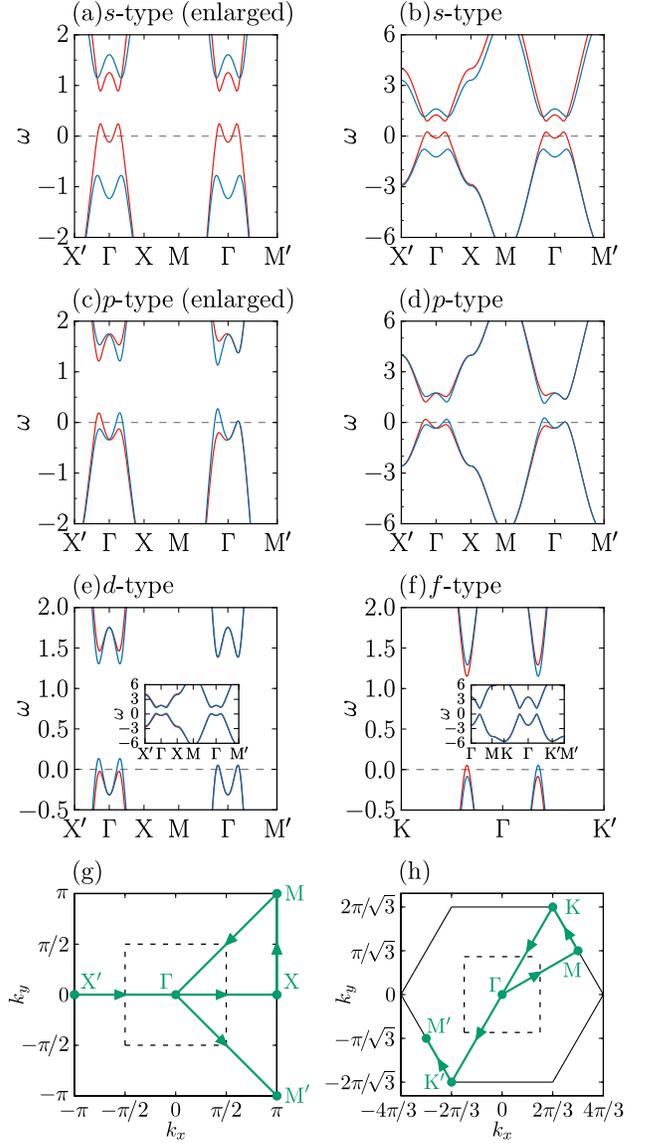}
\end{center}
\vspace*{-0.4cm} %add 2019/1/11
\caption{Calculated quasiparticle band dispersions in the spin-triplet excitonic phases of our model.  
Their energies $\omega$ are plotted in the Brillouin zone.  
The up-spin (red) and down-spin (blue) bands are illustrated.  We assume $N=1.94$ for the $s$-, $p$-, 
and $d$-type cross-hopping integrals and $N=1.98$ for the $f$-type cross-hopping integrals.  
We find $\phi=0$ for the $s$- and $d$-type, and $\phi=\pi/2$ for the $p$- and $f$-type.  
We set $D=6$, $U=9.5$, and $U^\prime=5$.  In (g) and (f), we show the Brillouin zones and the paths 
along which the band dispersions are drawn.  
}\label{fig:fig4}
\end{figure}

Let us first discuss the excitonic order parameter focusing on its phase, which can cause 
the $k$-space spin textures.  When the cross hopping is introduced, the phase of the excitonic 
order parameter is fixed to a certain value that depends on both the types of the cross hopping 
and electron filling $N$ \cite{KG16}.  The calculated results for the phase diagram are summarized 
in Fig.~\ref{fig:fig2}; the results in the parameter space of $(N,U')$ at $U=9$ are shown in 
Figs.~\ref{fig:fig2}(a)-(d) and the results in the parameter space of $(N,U)$ at $U'=5$ are shown 
in Figs.~\ref{fig:fig2}(e)-(h).  The calculated results for the amplitude $|\Phi_0^\mathrm{t}|$ and phase $\phi$ 
of the order parameter are also shown in Figs.~\ref{fig:fig2}(i)-(l) at $U = 9.5$ and $U'=5$ as 
a function of $N$.  In the normal phase where the order parameter is zero, the system is a band 
insulator at half filling ($N=2$) and a metal at $N<2$.  As the number of electrons decreases 
from $N=2$, the required $U$ for the excitonic phase transition increases.  In the excitonic 
phase, the phase of the order parameter strongly depends on the types of the cross hopping and 
$N$.  When the cross hopping is present, the $k$-space spin texture can emerge, where the 
splitting of the up-spin and down-spin bands occurs in the excitonic phase as is illustrated in 
Fig.~\ref{fig:fig3} and Fig.~\ref{fig:fig4}.  In the following, we discuss the $k$-space spin texture 
in the even-parity ($s$- and $d$-type) and odd-parity ($p$- and $f$-type) cross-hopping cases 
separately.  

% s and d type
For the $s$- and $d$-type cross-hopping integrals (even-parity case), we find that the phase 
is fixed to $\phi=\pi/2$ at $N=2$, which decreases with decreasing $N$ monotonically to zero.  
A finite value of the magnetization emerges, whose sign is opposite to the direction of 
the excitonic order parameter, in agreement with the preceding study \cite{KG16}.  
We define the magnetization of each orbital ($\ell = c,f$) as 
$m_0^{\ell}=\frac{1}{2L^2}\sum_{\bm{k},\sigma} \sigma \langle{\hat{n}}_{\bm{k},\sigma}^{\ell}\rangle$, 
which is calculated in the mean-field approximation 
${\varepsilon_{\ell}^\prime}(\bm{k}) \rightarrow {\varepsilon_{\ell}^\prime}(\bm{k})-U\sigma m_0^{\ell}$ 
and $\varepsilon_0 \rightarrow \varepsilon_0+U/2[ (m_0^c)^2+(m_0^f)^2 ]$.  
In the case of $s$-type cross hopping, there occurs the mixing between the orbital diagonal 
component of the order parameters (or magnetization) and the orbital off-diagonal component 
of the order parameter (or excitonic order), so that the excitonic order is accompanied necessarily 
by the magnetization.  From Eqs.~(\ref{eq:gamma^p}) and (\ref{eq:Epm}), we find that the quasiparticle 
band splits at $\phi \neq \pi/2$ since $h({\bm k})$ is real.  

The calculated $k$-space spin textures (or spin-dependent Fermi surfaces) are shown in 
Figs.~\ref{fig:fig3}(a) and \ref{fig:fig3}(c) for the $s$- and $d$-type even-parity cross-hopping 
integrals, respectively, and the corresponding quasiparticle band dispersions are shown in 
Figs.~\ref{fig:fig4}(a), \ref{fig:fig4}(b), and \ref{fig:fig4}(e).  
The time-reversal symmetry breaking by the excitonic ordering leads to 
${E}_{\bm{k},\sigma}^{\pm}={E}_{-\bm{k},\sigma}^{\pm}$ and ${E}_{\bm{k},\sigma}^{\pm}\neq{E}_{-\bm{k},-\sigma}^{\pm}$, 
whereby the degeneracy of the up-spin and down-spin bands is lifted along the $\omega$ direction.  
In the case of $s$-type cross-hopping integral, the excitonic order splits the up-spin and down-spin bands 
along the $\omega$ direction in the entire $k$-space, resulting in the net spin polarization.  
In the $d$-type cross-hopping integral, the excitonic order splits the spin bands as well, but due to 
the $k$-dependent spin occupation of the bands (or spin texture), the net spin polarization vanishes.  
Such a difference caused by the cross-hopping integrals affects the self-consistent equations, 
thereby giving rise to a qualitative difference in the orders of the excitonic phase transitions 
[see Figs.~\ref{fig:fig2}(i) and \ref{fig:fig2}(k)].  

% p and f type 
On the other hand, for the $p$- and $f$-type cross-hopping integrals (odd-parity case), 
we find that the phase is fixed to $\phi=0$ at $N=2$, increases continuously with decreasing $N$, 
and reaches a constant value $\pi/2$ at $N<1.96$ for the $p$-type and at $N<1.98$ for the $f$-type.  
We also find that the excitonic order parameter continuously decreases with decreasing $N$.  
Since $h({\bm k})$ is pure imaginary, the degeneracy of the up-spin and down-spin bands is 
lifted along the $k$ direction (rather than the $\omega$ direction) when $\phi \neq 0$.  
The $k$-space spin textures are shown in Figs.~\ref{fig:fig3}(b) and \ref{fig:fig3}(d) for the $p$- 
and $f$-type cross-hopping integrals, respectively, and the corresponding quasiparticle band 
dispersions are shown in Figs.~\ref{fig:fig4}(c), \ref{fig:fig4}(d), and \ref{fig:fig4}(f).  
The inversion symmetry breaking by the excitonic ordering leads to 
${E}_{\bm{k},\sigma}^{\pm}\neq{E}_{-\bm{k},\sigma}^{\pm} $ and 
${E}_{\bm{k},\sigma}^{\pm}={E}_{-\bm{k},-\sigma}^{\pm}$, whereby the splitting of the up-spin and 
down-spin bands emerges.  The splitting characteristic of the inversion symmetry breaking is 
clearly visible in the X$^\prime$--$\Gamma$--X line of the Brillouin zone [see Figs.~\ref{fig:fig4}(c) and \ref{fig:fig4}(d)] 
as well as in the K--$\Gamma$--K$^\prime$ line of the Brillouin zone [see Fig.~\ref{fig:fig4}(f)].  
 
%----------------------------------------------------------------------------------------%fig
%----------------------------------------------------------------------------------------%fig

%%%%%%%%%%%%%%%%%%%%%%%%%%%%%%%%
\subsection{Spin currents}
%%%%%%%%%%%%%%%%%%%%%%%%%%%%%%%%

Next, let us discuss the local and global spin currents in the excitonic phases of our model.  
The global spin current may be defined as 
\begin{align}
\bm{\hat{J}}_\mathrm{tot}^s 
=\! \frac{1}{L^2}\sum_{\bm{k},\sigma}\frac{\sigma}{2}
\begin{pmatrix}
 {\hat{c}}_{{\bm k},\sigma}^{\dagger} \! & \! {\hat{f}}_{{\bm k},\sigma}^{\dagger}
\end{pmatrix}
\bm{\nabla_k}
\left(\begin{array} {cc}
{\varepsilon}^\prime_{c}(\bm{k})  &  {\gamma}_\sigma^\prime(\bm{k})\\
{\gamma_\sigma^\prime}^{*}(\bm{k}) &{\varepsilon}^\prime_{f}(\bm{k})  
\end{array} \right)
\begin{pmatrix}
{\hat{c}}_{{\bm k},\sigma}^{\ }\\ {\hat{f}}_{{\bm k},\sigma}^{\ }
\end{pmatrix}
\label{eq:Jtot}
\end{align}
with $\bm{\nabla_k} =  \sum_{\tau}\bm{a}_{\tau}\pdv{}{k_\tau}$, which may be separated into 
the orbital diagonal component
\begin{align}
 \bm{\hat{J}}_{cc}^s + \bm{\hat{J}}_{ff}^s  =
&-\frac{1}{L^2}\sum_{\tau,\bm{k},\sigma} \sigma \, t_c \sin{k_\tau} {\hat{c}}_{{\bm k}, \sigma}^{\dagger} {\hat{c}}_{{\bm k}, \sigma}\bm{a}_\tau
\notag\\
&- \frac{1}{L^2}\sum_{\tau,\bm{k},\sigma} \sigma \, t_f \sin{k_\tau} {\hat{f}}_{{\bm k},\sigma}^{\dagger}{\hat{f}}_{{\bm k},\sigma}\bm{a}_\tau
\label{eq:Jccff}
\end{align}
and orbital off-diagonal component
\begin{align}
 \bm{\hat{J}}_{cf}^s + \bm{\hat{J}}_{fc}^s
&= \frac{1}{L^2}\sum_{\tau,\bm{k},\sigma} \sigma (-V_{\tau}\sin{k_\tau}+iV^\prime_{\tau}\cos{k_\tau})
{\hat{c}}_{{\bm k},\sigma}^{\dagger}{\hat{f}}_{{\bm k},\sigma} \bm{a}_\tau\notag\\
&+ \frac{1}{L^2}\sum_{\tau,\bm{k},\sigma} \sigma (-V_{\tau}\sin{k_\tau}-iV^\prime_{\tau}\cos{k_\tau})
{\hat{f}}_{{\bm k},\sigma}^{\dagger}{\hat{c}}_{{\bm k},\sigma}\bm{a}_\tau .
\label{eq:Jcffc}
\end{align}

\begin{figure}[tbh]
\begin{center}
\includegraphics[keepaspectratio=true,width=1.0\columnwidth]{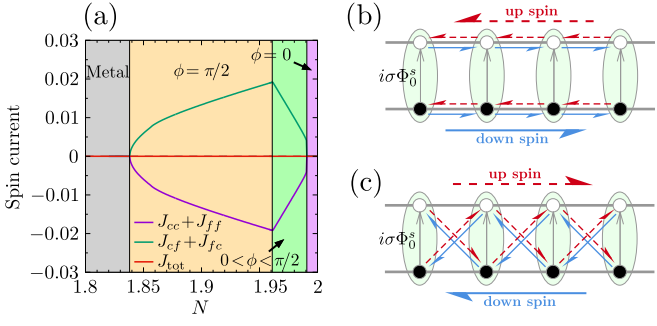}
\caption{(a) Calculated local and global spin currents as a function of the electron filling $N$, 
where we assume the $p$-type cross-hopping integral.  We set $D=6$, $U = 9.5$, and $U' = 5$.  
We also show schematic representations of the local spin currents in (b) and (c), where 
the diagonal and off-diagonal components are indicated, respectively. 
}\label{fig:fig5}
\end{center}
\end{figure}

Let us consider the diagonal component first.  
The expectation value of the diagonal component is given by 
\begin{align}
\langle \bm{\hat{J}}_{cc}^s \rangle + \langle \bm{\hat{J}}_{ff}^s \rangle =
-\frac{t}{L^2}\sum^{{\rm occ.}}_{\tau,\bm{k},\sigma} \sigma \frac{ \xi_{\bm k} 
\sin{k_\tau} }{\sqrt{ \xi_{\bm k}^2+|\gamma_\sigma^\prime(\bm{k})|^2}} \bm{a}_\tau ,
\label{eq:Jccff-exp}
\end{align}
where ${\rm occ.}$ means the summation over the $\bm{k}$ points at which the quasiparticle band $E^{-}_{\bm{k},\sigma}$ is occupied.  
In the even-parity case, we have $\xi_{\bm{k}} = \xi_{-\bm{k}}$ and 
$|{\gamma}_\sigma^\prime(\bm{k})|^2 = |{\gamma}_\sigma^\prime(-\bm{k})|^2$, 
which lead to $\langle \bm{\hat{J}}_{cc}^s \rangle + \langle \bm{\hat{J}}_{ff}^s \rangle =0$ 
since the integrand of Eq.~(\ref{eq:Jccff-exp}) becomes an odd function with respect to $\bm{k}$.  
Therefore, the diagonal component of the spin current never appears in the even-parity case.  
On the other hand, in the odd-parity case with $\phi \neq 0$ excitonic phases, we have 
$|{\gamma}_\sigma^\prime(\bm{k})|^2 \ne |{\gamma}_\sigma' (-\bm{k}) |^2$, which leads to 
$\langle \bm{\hat{J}}_{cc}^s \rangle + \langle \bm{\hat{J}}_{ff}^s \rangle \ne 0$.  
The diagonal component of the spin current calculated for the model with the $p$-type cross hopping 
is shown in Fig.~\ref{fig:fig5}, where the phase of the order parameter varies continuously from zero 
to $\pi/2$ with decreasing $N$ [see Fig.~\ref{fig:fig2}(j)].  
When $\phi = 0$, we have $|{\gamma}_\sigma' (\bm{k})|^2 = | {\gamma}'_\sigma (-\bm{k})|^2$, 
which leads to $\langle \bm{\hat{J}}_{cc}^s \rangle + \langle \bm{\hat{J}}_{ff}^s \rangle = 0$.  
However, when $\phi>0$, the diagonal component acquires a finite value, which is caused by 
the spin-dependent band splitting as shown in Fig.~\ref{fig:fig3}(b).  
The value of the diagonal component increases until the phase reaches $\pi/2$, but it decreases 
by further decreasing $N$ and vanishes when the excitonic order disappears.  

%----------------------------------------------------------------------------------------%fig
%----------------------------------------------------------------------------------------%fig

Let us next consider the off-diagonal component of the spin current.  
$V_\tau$ and $V'_\tau$ in Eq.~(\ref{eq:Jcffc}) depend on the types of the cross hopping: 
in the even-parity case, we have $V_\tau\neq0$ and $V^\prime_\tau=0$, and 
in the odd-parity case, we have $V_\tau=0$ and $V^\prime_\tau\neq0$.  
In the following, we consider the cases with the $s$- and $p$-type cross-hopping integrals.  

For the $s$-type cross-hopping integral, we have the off-diagonal component of the spin current as 
\begin{align}
 \langle \bm{\hat{J}}_{cf}^s \rangle + \langle \bm{\hat{J}}_{fc}^s \rangle
	= \frac{V}{L^2} \sum^{{\rm occ.}}_{\tau,\bm{k},\sigma} \sigma
		\frac{\Re [ \gamma_\sigma^\prime(\bm{k})]  \sin{k_\tau} }
			{\sqrt{ \xi_{\bm k}^2 + |\gamma_\sigma^\prime(\bm{k})|^2} }
		\bm{a}_\tau . 
\label{eq:Jcffc-exp-s}
\end{align}
Since ${\gamma}_\sigma^\prime(\bm{k}) = {\gamma}_\sigma^\prime(-\bm{k})$ in any $\phi$, 
the inversion symmetry remains in the excitonic phase.  
Therefore, the integrand of Eq.~(\ref{eq:Jcffc-exp-s}) is an odd function with respect to $\bm{k}$, 
so that $\langle \bm{\hat{J}}_{cf}^s \rangle + \langle \bm{\hat{J}}_{fc}^s \rangle =0$.  
The same discussion also applies to the case with the $d$-type cross-hopping integral.  

For the $p$-type cross-hopping integral, we have the off-diagonal component as 
\begin{align}
\langle \bm{\hat{J}}_{cf}^s \rangle + \langle \bm{\hat{J}}_{fc}^s \rangle
	= -\frac{V}{L^2} \sum^{{\rm occ.}}_{\tau,\bm{k},\sigma}
		\sigma \frac{\Im [ \gamma_\sigma^\prime(\bm{k})]  \cos{k_\tau} } 
	{\sqrt{ \xi_{\bm k}^2+|\gamma_\sigma^\prime(\bm{k})|^2} } \bm{a}_\tau .  
\label{eq:Jcffc-exp-p}
\end{align}
Since $|{\gamma}_\sigma^\prime(\bm{k})|^2 = | {\gamma}_\sigma^\prime(-\bm{k})|^2$ and 
$\Im[{\gamma}_\sigma^\prime(\bm{k})]= 2V\sum_{\tau} \sin{k_\tau}$ at $\phi = 0$, 
the integrand of Eq.~(\ref{eq:Jcffc-exp-p}) becomes an odd function with respect to $\bm{k}$, 
which leads to $\langle \bm{\hat{J}}_{cf}^s \rangle + \langle \bm{\hat{J}}_{fc}^s \rangle =0$.  
However, the inequality ${\gamma}_\sigma^\prime(\bm{k}) \neq {\gamma}_\sigma^\prime(-\bm{k})$ 
at $\phi=\pi/2$ breaks the inversion symmetry, which leads to 
$\langle \bm{\hat{J}}_{cf}^s \rangle + \langle \bm{\hat{J}}_{fc}^s \rangle \ne 0$.  
We obtain the same result whenever $\phi \neq 0$.  

The calculated off-diagonal component of the spin current for the $p$-type cross-hopping 
integrals is shown in Fig.~\ref{fig:fig5}.  One might think that this result contradicts the Bloch 
theorem that the current necessarily vanishes without external fields in the bulk systems.  
However, we find in Fig.~\ref{fig:fig5} that the sum of the diagonal and off-diagonal components 
exactly vanishes as 
\begin{align}
\langle\bm{\hat{J}}_{\mathrm{tot}}^s \rangle = \langle \bm{\hat{J}}_{cc}^s \rangle 
+ \langle \bm{\hat{J}}_{ff}^s \rangle 
+ \langle \bm{\hat{J}}_{cf}^s \rangle 
+ \langle \bm{\hat{J}}_{fc}^s \rangle = 0 , 
\label{eq:Jtot-sum}
\end{align}
which is consistent with the Bloch theorem.  
Thus, the global spin current always vanishes, in agreement with the identity at zero temperature \cite{VGKetal81,DT90}
\begin{align}
\langle\bm{\hat{J}}_{\mathrm{tot}}^s \rangle
=\frac{1}{L^2}\sum_{\bm{k},\sigma}^{\rm occ.} \sigma 
\bm{\nabla_k}E_{\bm{k},\sigma}^{-} = 0 . 
\label{eq:Jtot-band}
\end{align}

We should emphasis again that the global spin current never appears even if the cross-hopping 
integrals are added and the carriers are introduced.  The same discussion can also be applied to the 
case of the $f$-type cross-hopping integrals.  However, we note that the relation 
$\langle\bm{\hat{J}}_{\ell \ell'}^s \rangle=0$ necessarily holds since we have $\sum_{\tau}\bm{a}_{\tau}=0$ 
in the triangular lattice, which results in the vanishing local spin currents.  
As discussed in the study of superconductivity \cite{ohashi1996bloch, tada2016two} and also in a 
recent paper by Geffroy {\it et al.}~\cite{PhysRevB.97.155114}, the ground state containing a finite 
global current cannot be allowed in the equilibrium system.  In our mean-field calculation, we actually 
find that there is no global spin current but there can be a finite local spin current.  As shown here, 
the origin of the local spin current is the inversion-symmetry breaking in the excitonic phase.  
We again stress that the Bloch theorem does not prohibit the presence of the local spin currents.  
The detailed discussions on the absence of the global spin current and the presence of the local 
spin currents are found in Appendices A and B.  

\section{Summary}

We studied the $k$-space spin textures and spin currents in the spin-triplet excitonic phase of 
the two-band Hubbard model defined on the square and triangular lattices by the mean-field 
approximation.  We assumed the noninteracting band structure with a direct band gap and 
introduced the $s$-, $p$-, $d$-, and $f$-type cross-hopping integrals.  We thus found that, 
depending on the types of the cross hopping, interaction strength, and electron filling, the phase 
of the excitonic order parameter is fixed to be imaginary, whereby the $k$-space spin texture 
and local spin current can emerge.  

The even-parity cross-hopping integrals of the $s$- and $d$-type lift the spin degeneracy of 
the band dispersions by the breaking of the time-reversal symmetry, which leads to the 
$k$-space spin texture, whereas the local spin current exactly vanishes because the 
space-inversion symmetry remains in this system.  
On the other hand, the odd-parity cross-hopping integrals of the $p$- and $f$-type lift the 
spin degeneracy of the band structures by the breaking of the space-inversion symmetry, 
which leads to the $k$-space spin texture as well.  Moreover, in the case of the $p$-type 
cross-hopping integral, the local spin currents of the diagonal and off-diagonal components 
remain finite when the excitonic order parameter has the imaginary value.  The global spin 
current always vanishes, which is consistent with the Bloch theorem.  

The experimental observation of the $k$-space spin textures and local spin currents may, 
therefore, be very useful for verification of the presence of the spin-triplet excitonic orders.  
We hope that our results will encourage experimental confirmations of the excitonic phases 
in real materials.  

\begin{acknowledgements} 
We thank Prof.~Yoji Ohashi for tutorial lectures and enlightening discussions.  
This work was supported by Grants-in-Aid for Scientific Research 
(Nos.~JP26400349, JP15H06093, JP17K05530, and JP18K13509) from JSPS of Japan.  
The numerical calculations were carried out on computers at Yukawa Institute 
for Theoretical Physics, Kyoto University, Japan.  
\end{acknowledgements}

\appendix

\section{Absence of the global spin current}

Here, we discuss the absence of the global spin current in spin-triplet excitonic insulator states 
from the viewpoint of the Bloch theorem.  First, we introduce the spin current from the continuity 
equation in the $d$-dimensional lattice containing $L^d$ sites and define the global and local 
(or partial) spin currents.  Next, we derive the Bloch theorem for the global spin current and 
examine its correspondence with the results of our mean-field calculations.  

\subsection{Global and local spin currents}

The current operators may be derived from the continuity conditions of the Hamiltonian as 
\cite{mahan1990} 
\begin{align}
\frac{\partial}{\partial t}\hat{S}^{z}_{j}
=i[\hat{\mathcal{H}},\hat{S}^{z}_{j}]=
-\sum_{\tau}(\hat{J}^{z}_{(j,\tau)}-\hat{J}^{z}_{(j,-\tau)}), 
\end{align}
where $\hat{J}^{z}_{(j,\tau)}$ denotes the operator of the spin current flowing out from site $j$ to 
$j+\tau$ and $\hat{S}^{z}_{j}=1/2\sum_{\sigma}\sigma (\hat{n}^{c}_{j\sigma}+\hat{n}^{f}_{j\sigma})$.  
We note that this argument is justified if and only if the system has the axial spin rotational 
symmetry about the $z$-axis, $R_{z}$, and the expectation value of the $z$-component of the 
total spin operator is conserved.  Thus, we assume in the following discussions that the system 
has the symmetry $R_{z}$.  The spin current operators $\hat{J}^{z}_{(j,\tau)}$ may then be divided 
into the orbital diagonal and orbital off-diagonal components as 
\begin{align}
&\hat{J}^{z}_{(j,\tau),cc}
=-it_{c}/2\sum_{\sigma}\sigma(
\hat{c}^{\dagger}_{j+\tau,\sigma}\hat{c}_{j,\sigma}-\mathrm{H.c.}),\\
& \hat{J}^{z}_{(j,\tau),ff}
=-it_{f}/2\sum_{\sigma}\sigma(
\hat{f}^{\dagger}_{j+\tau,\sigma}\hat{f}_{j,\sigma}-\mathrm{H.c.}),\\
&\hat{J}^{z}_{(j,\tau),cf}
=-iV_{1,\tau}/2\sum_{\sigma}\sigma(
\hat{c}^{\dagger}_{j+\tau,\sigma}\hat{f}_{j,\sigma}-\mathrm{H.c.}),\\
&\hat{J}^{z}_{(j,\tau),fc}
=-iV_{2,\tau}/2\sum_{\sigma}\sigma(
\hat{f}^{\dagger}_{j+\tau,\sigma}\hat{c}_{j,\sigma}-\mathrm{H.c.}).
\end{align}
Then, the global spin current flowing in the direction $\tau$ may be defined as 
\begin{align}
L^{d}\hat{J}^{z}_{\tau}=\sum_{j}\hat{J}^{z}_{(j,\tau)}
=\frac{1}{2}\sum_{\bm{k}\sigma}\sigma
\mqty(\hat{c}^{\dagger}_{\bm{k},\sigma}~\hat{f}^{\dagger}_{\bm{k},\sigma})
\frac{\partial \mathcal{H}(\bm{k})}{\partial k_{\tau}}
\mqty(\hat{c}_{\bm{k},\sigma}\\\hat{f}_{\bm{k},\sigma})
\label{eq:global_spin_current}
\end{align}
with 
\begin{align}
\mathcal{H}(\bm{k})=
\mqty(\varepsilon_{c}(\bm{k})&\gamma(\bm{k})\\
\gamma^{*}(\bm{k})&\varepsilon_{f}(\bm{k})), 
\end{align}
where the $\tau$ denotes the vector from site $j$ to site $j+\tau$.  Note that the 
interaction terms of the Hamiltonian Eq.~(\ref{eq:Hint}) do not contribute to the spin 
current operators.  

Similarly, the partial spin current may be defined as 
\begin{align}
L^{d}\hat{J}^{z}_{\tau,ll'}=\sum_{j}\hat{J}^{z}_{(j,\tau),ll'} , 
\end{align}
where $l$ and $l'$ $(=c,f)$ denote the orbitals.  
We also define the difference between the orbital diagonal and orbital off-diagonal 
spin currents as 
\begin{align}
\hat{J}^{\prime z}_{\tau}=&\hat{J}^{z}_{\tau,cc}
+\hat{J}^{z}_{\tau,ff}-\hat{J}^{z}_{\tau,cf}-\hat{J}^{z}_{\tau,fc}
\nonumber\\
=&\frac{1}{2L^d}\sum_{\bm{k}\sigma}\sigma
\mqty(\hat{c}^{\dagger}_{\bm{k},\sigma}~\hat{f}^{\dagger}_{\bm{k},\sigma})\tau_{z}
\frac{\partial \mathcal{H}(\bm{k})}{\partial k_{\tau}}\tau_{z}
\mqty(\hat{c}_{\bm{k},\sigma}\\\hat{f}_{\bm{k},\sigma}) , 
\label{eq:partial_spin_current}
\end{align}
where $\tau^{z}$ denotes the $z$-component of the Pauli matrix.  
We may then obtain the orbital diagonal and orbital off-diagonal spin currents as 
$\hat{J}^{(+)}_{\tau}=\hat{J}^{z}_{\tau}+\hat{J}^{\prime z}_{\tau}$ and 
$\hat{J}^{(-)}_{\tau}=\hat{J}^{z}_{\tau}-\hat{J}^{\prime z}_{\tau}$, respectively.  

Here, we note that the orbital-decomposed partial spin currents may be termed as 
the {\em local} spin currents if the orbitals are located in different spatial positions, as 
is assumed in the main text.  
The global (or total) spin current may then be defined as a sum of the local (or partial) spin 
currents.  We also note that the global spin current if exists may obviously be observed 
experimentally but the local spin currents should in principle be observed experimentally 
as well, which can lead to an experimental proof of the existence of the spin-triplet 
excitonic insulator state.  
In the Appendices A and B, we use the term ``partial'' spin current rather than ``local'' 
spin current.  

\subsection{The Bloch theorem}

Now, let us prove the Bloch theorem for our system, which states that the persistent 
spin current does not exist in thermal equilibrium without any external fields.  
The proof is carried out in the following two steps \cite{tada2016two}.  
First, we introduce the excited state generated by an infinitesimal twisting of the 
spin-dependent Peierls phase in the hopping parameters.  
Secondly, using the inequality originated from the passivity (defined below) of thermal 
equilibrium states, we show on the basis of the dimensional analysis that a contradiction 
is lead if we assume the existence of the global spin current.  In this proof, we assume 
that the system is under the periodic boundary condition in all the orthogonal directions.  

First, we introduce the spin-dependent Peierls phase using the twist operator defined as 
\begin{align}
\hat{U}(\bm{\varphi})=\exp(i\bm{\varphi}\cdot\sum_{l,j,\sigma}\sigma \hat{n}^{l}_{j,\sigma}\bm{r}_{j}), 
\end{align}
where $\bm{\varphi}$ denotes a vector in the reciprocal lattice space, which satisfies 
$L\bm{\varphi}\cdot\bm{a}_{i}=0$ and its amplitude characterizes the intensity
of the flux penetrating a one-dimensional ring.  Thus, the vector $\bm{\varphi}$ can be
written as
\begin{align}
\bm{\varphi}=\frac{1}{L}\sum_{j}m_{j}\bm{b}_{j}=\mathcal{O}(L^{-1}),~~m_{j}\in\mathbb{Z}.
\end{align}
where $\bm{a}_{i}\cdot\bm{b}_{j}=2\pi\delta_{ij}$.
Here, we assume the integers $m_{j}\in\mathbb{Z}$ are sufficiently smaller than $L$, so that 
the vector $\bm{\varphi}$ has the order of $L^{-1}$.
Using this twist operator, the fermion creation and annihilation operators with momentum 
$\bm{k}$ are transformed into the other fermion operators with momentum 
$\bm{k}-\sigma\bm{\varphi}$ as
\begin{align}
&\hat{U}^{\dagger}(\bm{\varphi})\hat{c}_{\bm{k},\sigma}\hat{U}(\bm{\varphi})
=\hat{c}_{\bm{k}-\sigma\bm{\varphi},\sigma},
\\
&\hat{U}^{\dagger}(\bm{\varphi})\hat{c}^{\dagger}_{\bm{k},\sigma}\hat{U}(\bm{\varphi})
=\hat{c}^{\dagger}_{\bm{k}-\sigma\bm{\varphi},\sigma},
\\
&\hat{U}^{\dagger}(\bm{\varphi})\hat{f}_{\bm{k},\sigma}\hat{U}(\bm{\varphi})
=\hat{f}_{\bm{k}-\sigma\bm{\varphi},\sigma},
\\
&\hat{U}^{\dagger}(\bm{\varphi})\hat{f}^{\dagger}_{\bm{k},\sigma}\hat{U}(\bm{\varphi})
=\hat{f}^{\dagger}_{\bm{k}-\sigma\bm{\varphi},\sigma}, 
\end{align}
where we note that the shifted momentum $\bm{k}-\sigma\bm{\varphi}$ is in the
Brillouin zone.  Because thermal equilibrium states are passive (or energetically stable) 
for any local unitary transformation \cite{Bratteli1981, tada2016two}, we can introduce 
the following inequality:
\begin{align}
\omega_{0}(\hat{U}^{\dagger}(\bm{\varphi})[\hat{\mathcal{H}},\hat{U}(\bm{\varphi})])\ge 0 , 
\label{eq:passivity.condition}
\end{align}
where $\omega_{0}(\cdots)$ is defined as the expectation value with respect to the 
infinite thermodynamical equilibrium state.  In particular, if the $N$-fermion system has 
a unique ground state $\ket{\Phi^{(N)}_{0}}$ at zero temperature, $\omega_{0}(\cdots)$ may 
be rewritten as 
\begin{align}
\omega_{0}(\cdots)=\lim_{L\rightarrow\infty}\bra{\Phi^{(N)}_{0}}\cdots\ket{\Phi^{(N)}_{0}} , 
\end{align}
where we take the infinite volume limit $L\rightarrow\infty$ so that the density 
$\rho=N/L^{d}$ converges to a finite positive constant.  

Using the twist operator, we then obtain
\begin{align}
\hat{U}^{\dagger}(\bm{\varphi})[\hat{\mathcal{H}},\hat{U}(\bm{\varphi})] = &\,
2\bm{\varphi}\cdot\sum_{\bm{k}\sigma}\frac{\sigma}{2}
\mqty(\hat{c}^{\dagger}_{\bm{k},\sigma}~\hat{f}^{\dagger}_{\bm{k},\sigma})
\nabla_{\bm{k}}\mathcal{H}({\bm k}) 
\mqty(\hat{c}_{\bm{k},\sigma}\\\hat{f}_{\bm{k},\sigma})
\nonumber\\&
+\mathcal{O}(L^{d-2}) , 
\end{align}
where the first term of the right-hand side corresponds to the global spin current 
defined in Eq.~(\ref{eq:global_spin_current}).  If the system has a nonzero bulk spin current, 
this term is of the order of $L^{d-1}$. Thus, we find 
\begin{align}\frac{1}{L^{d}}\omega_{0}(\hat{U}^{\dagger}(\bm{\varphi})
[\hat{\mathcal{H}},\hat{U}(\bm{\varphi})])
=&\,\frac{4\pi}{L}\sum_{\tau}m_{\tau}\omega_{0}(\hat{J}^{z}_{\tau})
\nonumber\\&+\mathcal{O}(L^{-2})\ge 0 , 
\label{eq:BlochTheorem}
\end{align}
where we note that the vector $\bm{\varphi}$ is arbitrary, so that we can take any values 
of $m_\tau$.  Now, if we assume the presence of the positive global spin current, i.e., 
$\omega(\hat{J}^z_{\tau})> 0$, then choosing all $m_{\tau}$ to be negative, we obtain 
$\sum_{\tau}m_{\tau}\omega_{0}(\hat{J}_{\tau})< 0$, which contradicts 
the passivity condition Eq.~(\ref{eq:passivity.condition}).  
Therefore, we find that the global spin current does not exist.  
In other words, the axial spin rotational symmetry about the $z$-axis is not broken 
in the ground state of the system.  
We also note that the above argument cannot be applied to the case of the surface 
currents.  If the system has only the surface currents, the leading order of 
Eq.~(\ref{eq:BlochTheorem}) becomes $L^{-2}$.  Therefore, the Bloch theorem does not 
prohibit the existence of the surface currents.  Similarly, the bulk spin current is robust 
against surface defects because of the same reasons.  
As discussed in Appendix B, such a dimensional analysis can also be applied to the 
proof of the existence of the partial spin currents, which are not prohibited by the 
Bloch-like theorem in general.  

\subsection{Absence of the global spin current in the mean-field approximation}

Here, we discuss the validity of the Bloch theorem in the straightforward mean-field calculation.  
In general, the Bloch theorem is applicable to any interacting electron systems with the axial 
spin rotational symmetry and therefore should be valid in the mean-field approximation as well.  
Using the Hellmann-Feynman theorem \cite{martin2004electronic}, we obtain 
\begin{align}
\langle\hat{\bm{J}}_{\mathrm{tot}}^s \rangle
=&\frac{1}{L^d}\sum_{\tau,\bm{k},\sigma} \sigma 
\left(\pdv{E_{\bm{k},\sigma}^{+}}{k_\tau} f(E_{\bm{k},\sigma}^{+})
+\pdv{E_{\bm{k},\sigma}^{-}}{k_\tau}f(E_{\bm{k},\sigma}^{-}) \right)\bm{a}_\tau
\notag\\
=&\frac{1}{L^d}\sum_{\bm{k},\sigma,\nu=\pm} \sigma 
(\bm{\nabla_k}E_{\bm{k},\sigma}^{\nu})  f(E_{\bm{k},\sigma}^{\nu})
\label{eq:app-Jtot}
\end{align}
where $\nu$ denotes the band index.  Using the density of states defined as 
\begin{align}
D_{\nu,\sigma}(E)dE= \frac{L^d}{(2\pi)^d} \left[\int_{E_{\bm{k},\sigma}^{\nu}=E}
\frac{ dl }{ \qty| \bm{\nabla_k}E_{\bm{k},\sigma}^{\nu} | } \right] dE ,
\end{align}
where $dl$ is the surface element in $\bm{k}$-space satisfying $E_{\bm{k},\sigma}^\nu = E$, 
we can rewrite Eq.~(\ref{eq:app-Jtot}) as 
\begin{align}
\langle\bm{\hat{J}}_{\mathrm{tot}}^s \rangle
=&\frac{1}{L^d}\sum_{\sigma,\nu}\sigma \int  D_{\nu,\sigma}(E_{\sigma}^{\nu})
(\bm{\nabla_k}E_{\bm{k},\sigma}^{\nu}) f(E_{\sigma}^{\nu}) dE_{\sigma}^{\nu} \notag\\
=&\frac{1}{(2\pi)^d}\sum_{\sigma,\nu}\sigma \int dE_{\sigma}^{\nu} f(E_{\sigma}^{\nu}) 
\int \bm{n}dl =0
\end{align} 
with $\bm{n}=\nabla_{\bm{k}}E^{\nu}_{\bm{k},\sigma}/\qty|\nabla_{\bm{k}}E^{\nu}_{\bm{k},\sigma}|$.  
The integral over the closed constant-energy surface vanishes $\int \bm{n}dl=0$, resulting in 
the vanishing global spin current.  

\section{Existence of the partial spin current}

Here, we discuss the existence of the partial spin current in spin-triplet excitonic insulator 
states.  First, we make the Bloch-like argument for the partial spin current [defined in 
Eq.~(\ref{eq:partial_spin_current})] as an application of the method given in Appendix A.  
Next, we make the argument based on the discrete lattice symmetries.  

\subsection{Argument based on the Bloch-like theorem}

Introducing the operator defined as
\begin{align}
\hat{W}=&\exp{i\frac{\pi}{2}\sum_{j,\sigma}
\mqty(\hat{c}^{\dagger}_{j,\sigma}~\hat{f}^{\dagger}_{j,\sigma})
\tau^{z}\mqty(\hat{c}_{j,\sigma}\\\hat{f}_{j,\sigma})}, 
\end{align}
we find that the fermion creation and annihilation operators for the $c$- and $f$-band 
electrons are transformed as 
\begin{align}
&\hat{W}^{\dagger}\hat{c}_{j,\sigma}\hat{W}=i\hat{c}_{j,\sigma} ,
\\
&\hat{W}^{\dagger}\hat{c}^{\dagger}_{j,\sigma}\hat{W}=-i\hat{c}^{\dagger}_{j,\sigma} ,
\\
&\hat{W}^{\dagger}\hat{f}_{j,\sigma}\hat{W}=-i\hat{f}_{j,\sigma} ,
\\
&\hat{W}^{\dagger}\hat{f}^{\dagger}_{j,\sigma}\hat{W}=i\hat{f}^{\dagger}_{j,\sigma}.
\end{align}
Thus, using this operator $\hat{W}$ and the twist operator defined in Eq.~(\ref{eq:global_spin_current}), 
we obtain
\begin{align}
&\frac{1}{L^d}(\hat{U}(\bm{\varphi})\hat{W})^{\dagger}[\hat{\mathcal{H}},(\hat{U}(\bm{\varphi})\hat{W})]
\nonumber\\
&~~~=\frac{1}{L^{d}}\sum_{\bm{k},\sigma}\qty(\hat{c}^{\dagger}_{\bm{k},\sigma}~
\hat{f}^{\dagger}_{\bm{k}\sigma})\tau^{z}[\mathcal{H}_{\sigma}(\bm{k}),\tau^{z}]
\mqty(\hat{c}_{\bm{k},\sigma}\\\hat{f}_{\bm{k},\sigma})
\nonumber\\
&~~~~~~+2\bm{\varphi}\cdot\hat{\bm{J}}^{\prime z}+\mathcal{O}(L^{-2}), 
\end{align}
where we note that the first and second terms of the right-hand side are of the orders 1 
and $L^{-1}$, respectively, if we assume that the bulk partial spin current exists.  
However, unless the commutator $[\mathcal{H}_{\sigma}(\bm{k}),\tau_{z}]$ is zero, 
the straightforward Bloch-like argument cannot be applied to the present case.  
In other words, because the first term is larger than the second one, we do not 
obtain the contradiction to the passivity of the thermal equilibrium states.  
Thus, in general, the partial spin current is not prohibited by the Bloch-like argument 
as long as there is the interorbital hybridization satisfying $[\hat{W},\hat{\mathcal{H}}]\ne 0$.  
In this sense, vanishing of the expectation value of the first term is a sufficient 
condition to prohibit the partial spin current.  
It should be noted that this condition is already broken in systems with the 
cross-hopping terms.  However, the partial spin current does not appear in the 
normal phases,  which is due to the other conditions associated with the lattice 
symmetries.  As discussed below, the partial spin current emerges as a result 
of the ``inversion'' symmetry breaking in the excitonic phases.  

\subsection{Argument based on the discrete lattice symmetries}
 
Now, let us prove the existence of the partial spin current from the viewpoint 
of the symmetries that are  broken in the excitonic phases.  Our strategy is 
based on the following two assumptions:  
(i) The ground state of our system is unique.  
(ii) There is no symmetry operation $\hat{g}$ that anticommutes with the current operator $\hat{J}_{\tau}$.  
These two assumptions are naturally applicable to our mean-field solutions obtained as the stationary points 
of the free energy.  
The relevance of these assumption may be confirmed as follows:  
If there is at least one symmetry operation $\hat{g}$ that anticommutes with the current operator 
$\{\hat{g},\hat{J}_{\tau}\}=0$, we obtain
\begin{align}
0=\bra{\psi}\hat{g}^{-1}
\{\hat{J}_{\tau},\hat{g}\}\ket{\psi}
=&2\bra{\psi}\hat{J}_{\tau}\ket{\psi} , 
\end{align}
where we use the uniqueness of the ground state, i.e., 
$\hat{g}\ket{\Psi}=e^{i\theta}\ket{\Psi}$ except for an arbitrary phase $\theta$.  
Thus, we find that the partial spin current is absent as long as the conditions (i) and (ii) 
are satisfied.

Next, let us examine the symmetries of our two-band Hubbard model.  
For simplicity, we consider the symmetries of the one body part of the Hamiltonian 
only and treat the interaction terms within the mean-field approximation.  It is, however, 
not difficult to extend our argument to the interacting systems.  
As discussed in the main text, we have two types of the cross-hopping integrals, i.e., 
either with \textit{even} parity ($s$-type) or with \textit{odd} parity ($p$-type), where 
the latter has a sign change $\bm{k}\rightarrow-\bm{k}$ for the spatial inversion.  
The mean-field Hamiltonian transforms under the time-reversal symmetry operation ($\mathcal{T}$) 
or under the space-inversion symmetry operation ($\mathcal{P}$) as 
\begin{align}
&\mathcal{T}\mathcal{H}(\bm{k})\mathcal{T}^{-1}=\mathcal{H}(-\bm{k}) ,
\label{eq:time.reversal.symmetry}
\\
&\mathcal{P}\mathcal{H}(\bm{k})\mathcal{P}^{-1}=\mathcal{H}(-\bm{k}) , 
\label{eq:space.inversion.symmetry}
\end{align}
where we note that $\mathcal{P}$ is a unitary operator satisfying $\mathcal{P}^2=1$ while 
$\mathcal{T}$ is an antiunitary operator containing the complex conjugate operation 
$\mathcal{K}$.\footnote{Note that we use a general definition Eq.~(\ref{eq:time.reversal.symmetry}) 
of the time-reversal symmetry operation in this Appendix; another definition, which uses the 
antiunitary operator that changes the signs of $\bm{k}$ and spin, does not satisfy the condition (ii).}  
In our system, there are several candidates for these symmetry operations, which depend on both 
the parity of the cross-hopping term $l\in\{s,p\}$ and the phase of the excitonic order parameter 
$\phi\in\{0,\frac{\pi}{2}\}$.  To see this explicitly, it is instructive to rewrite our mean-field Hamiltonian 
using the Pauli matrices as
\begin{align}
&\mathcal{H}_\mathrm{MF}(\bm{k})=
\varepsilon^{\prime}_{+}(\bm{k})~I_{2}\otimes I_{2}
+\varepsilon^{\prime}_{-}(\bm{k})~\tau^{z}\otimes I_{2}
\nonumber\\&+\left\{
\begin{array}{cc}
\gamma_{+}(\bm{k})~\tau^{x}\otimes I_{2}
+\Delta_{0}~\tau^{x}\otimes\sigma^{z}
,&(l,\phi)=(s,0)\\
\gamma_{+}(\bm{k})~\tau^{x}\otimes I_{2}
+\Delta_{\pi}~\tau^{y}\otimes\sigma^{z}
,&(l,\phi)=(s,\frac{\pi}{2})\\
\gamma_{-}(\bm{k})~\tau^{y}\otimes I_{2}
+\Delta_{0}~\tau^{x}\otimes\sigma^{z}
,&(l,\phi)=(p,0)\\
\gamma_{-}(\bm{k})~\tau^{y}\otimes I_{2}
+\Delta_{\pi}~\tau^{y}\otimes\sigma^{z}
,&(l,\phi)=(p,\frac{\pi}{2})
\end{array}
\right.
\end{align}
where
\begin{align}
&2\varepsilon^{\prime}_{\pm}(\bm{k})=\varepsilon^{\prime}_{c}(\bm{k})\pm\varepsilon^{\prime}_{f}(\bm{k}),\\
&\gamma_+(\bm{k})=2\sum_{\tau}V_{\tau}\cos{k_{\tau}},\\
&\gamma_-(\bm{k})=2\sum_{\tau}V_{\tau}\sin{k_{\tau}},\\
&2\Delta_{\phi}=-U^{\prime}\left|\Phi^{t}_{0}\right|e^{-i\phi} ,
\end{align}
and $\otimes$ denotes the tensor product of two matrices.  
$\tau^{\alpha}$ and $\sigma^{\alpha}$ denote the Pauli matrices for the orbital and spin degrees 
of freedom, respectively.  We note that $\gamma_{-}(\bm{k})$ is an odd function with respect to 
the inversion $\bm{k}\rightarrow-\bm{k}$.  Then, if we note the relations 
$\{\sigma^{a},\sigma^{b}\}=\{\tau^{a},\tau^{b}\}=2\delta_{ab}$ and 
$\{\tau^{y},\mathcal{K}\}=\{\sigma^{y},\mathcal{K}\}=0$, 
we can easily write down the time-reversal $\mathcal{T}$ and space inversion $\mathcal{P}$ 
symmetries such that Eqs.~(\ref{eq:time.reversal.symmetry}) and (\ref{eq:space.inversion.symmetry}) 
are satisfied.  In fact, we can choose $\mathcal{T}=I_{2}\otimes\sigma^{x}\mathcal{K}$ and 
$\mathcal{P}=\tau^{z}\otimes\sigma^{x}$ for $(l,\phi)=(p,\frac{\pi}{2})$.  
For other $(l,\phi)$, we can also choose $\mathcal{T}$ and $\mathcal{P}$ in the same manner.  

Then, let us examine whether our mean-field Hamiltonian has the time-reversal $\mathcal{T}$ 
or space-inversion $\mathcal{P}$ symmetry that satisfies the condition (ii) given above.  
Using the mean-field Hamiltonian, the global and partial spin currents in Eqs.~(\ref{eq:global_spin_current}) 
and (\ref{eq:partial_spin_current}) can be rewritten, respectively, as 
\begin{align}
\mathcal{J}^{z}_{\tau}(\bm{k})=&~
\partial_{\tau}\varepsilon^{\prime}_{+}(\bm{k})~
I_{2}\otimes \sigma^{z}
+\partial_{\tau}\varepsilon^{\prime}_{-}(\bm{k})~
\tau^{z}\otimes \sigma^{z}
\nonumber\\&+\left\{
\begin{array}{cc}
\partial_{\tau}\gamma_{+}(\bm{k})\tau^{x}~
\otimes\sigma^{z}&(l=s)\\
\partial_{\tau}\gamma_{-}(\bm{k})\tau^{y}~
\otimes\sigma^{z}&(l=p)
\end{array}
\right. ,\\
\mathcal{J}^{\prime z}_{\tau}(\bm{k})=&~
\partial_{\tau}\varepsilon^{\prime}_{+}(\bm{k})~
I_{2}\otimes \sigma^{z}
+\partial_{\tau}\varepsilon^{\prime}_{-}(\bm{k})~
\tau^{z}\otimes \sigma^{z}
\nonumber\\&-\left\{
\begin{array}{cc}
\partial_{\tau}\gamma_{+}(\bm{k})~
\tau^{x}\otimes\sigma^{z}&(l=s)\\
\partial_{\tau}\gamma_{-}(\bm{k})~
\tau^{y}\otimes\sigma^{z}&(l=p)
\end{array}
\right. ,
\end{align}
where we use the following notations: 
\begin{align}
\hat{O}=&\sum_{\bm{k}}\hat{\bm{c}}^{\dagger}_{\bm{k}}
\mathcal{O}(\bm{k})\hat{\bm{c}}_{\bm{k}},\\ 
\hat{\bm{c}}_{\bm{k}}=&
(\hat{c}_{\bm{k},\uparrow}
~\hat{f}_{\bm{k},\uparrow}
~\hat{c}_{\bm{k},\downarrow}
~\hat{f}_{\bm{k},\downarrow})^{T},\\
\hat{\bm{c}}^{\dagger}_{\bm{k}}=&
(\hat{c}^{\dagger}_{\bm{k},\uparrow}
~\hat{f}^{\dagger}_{\bm{k},\uparrow}
~\hat{c}^{\dagger}_{\bm{k},\downarrow}
~\hat{f}^{\dagger}_{\bm{k},\downarrow}).
\end{align}
By a straightforward calculation, we obtain the time-reversal symmetry $\mathcal{T}$ and 
space-inversion symmetry $\mathcal{P}$ that satisfy the relations 
$\mathcal{T}\mathcal{J}^{\prime z}_{\tau}(\bm{k})
\mathcal{T}^{-1}=-\mathcal{J}^{\prime z}_{\tau}(-\bm{k})$ 
and 
$\mathcal{P}\mathcal{J}^{\prime z}_{\tau}(\bm{k})
\mathcal{P}^{-1}=-\mathcal{J}^{\prime z}_{\tau}(-\bm{k})$,
which correspond to the anticommutation relation for the partial spin current operator 
$\hat{J}^{\prime z}_{\tau}$, as shown in Table~\ref{Table:symmetryofspincurrents}.

% TABLE I

\begin{table}[h!]
\caption{
The time-reversal and space-inversion symmetries for the mean-field Hamiltonian
with the indexes $(l,\phi)$.  $\omega$ and $\omega\mathcal{K}$ (`--') denote the system 
with (without) time-reversal symmetry $\mathcal{T}=\omega\mathcal{K}$ or 
space-inversion symmetry $\mathcal{P}=\omega$.  Here, $\omega$ is the unitary matrix defined as 
$\omega=e^{i\phi I_{2}\otimes\sigma^{z}}=\cos\phi~I_{2}\otimes I_{2}+i\sin\phi~I_{2}\otimes \sigma^{z}$.
Note that the other time-reversal or space-inversion symmetries, which anticommute with the spin 
current operators, do not exist except for $\mathcal{T}=\omega\mathcal{K}$ and $\mathcal{P}=\omega$.
}
\begin{tabular}{|l|c|c|c|c|}
\hline
&\multicolumn{2}{|c|}{global}
&\multicolumn{2}{|c|}{partial}
\\ \cline{2-5}
&$\mathcal{T}$&$\mathcal{P}$
&$\mathcal{T}$&$\mathcal{P}$\\
\hline\hline
$(l,\phi)=(s,0)$
&$\omega\mathcal{K}$
&$\omega$
&$\omega\mathcal{K}$
&$\omega$
\\
\hline
$(l,\phi)=(s,\frac{\pi}{2})$
&--
&$\omega$
&--
&$\omega$
\\
\hline
$(l,\phi)=(p,0)$
&$\omega\mathcal{K}$
&--
&$\omega\mathcal{K}$
&--
\\
\hline
$(l,\phi)=(p,\frac{\pi}{2})$
&--
&--
&--
&--
\\
\hline
\end{tabular}
\label{Table:symmetryofspincurrents}
\end{table}

In particular, if $(l,\phi)=(p,\pi/2)$, we find that there is no corresponding time-reversal 
$\mathcal{T}$ or space-inversion $\mathcal{P}$ symmetries in the system.  
In other words, because the mean-field Hamiltonian does not satisfy the condition (ii), 
the global and partial spin currents are allowed by the symmetries $\mathcal{T}$ and $\mathcal{P}$.  
However, the global spin current is prohibited by the Bloch theorem, so that only the partial spin 
current is allowed.  Moreover, such symmetry breakings in the excitonic phase may lead to the 
asymmetry of the band structures, resulting in the $k$-space spin textures.  
In this sense, the partial spin current is a signature of the absence of the time-reversal $\mathcal{T}$ and 
space-inversion $\mathcal{P}$ symmetries in the system.  
It should be noted, however, that we do not deny the possible existence of the other symmetries 
that satisfy the conditions (i) and (ii).  In fact, the model with the $f$-type cross hopping has the 
3-fold rotational symmetry $C_{3}$.  Then, even if both the time-reversal $\mathcal{T}$ and 
space-inversion $\mathcal{P}$ symmetries are broken in the excitonic phase, the partial spin 
currents are canceled out due to the $C_{3}$ symmetry.  

Finally, let us make a remark on our derivation of the partial spin currents.  
In this Appendix, we use two approaches to prove the existence of the partial spin current.  
However, we should note that the arguments given in both of these two approaches are not 
the necessary condition, but they are the sufficient condition for the absence of the partial 
spin current.  
In other words, the existence of the partial spin currents is allowed only if the system has 
the cross hopping satisfying $[\hat{W}, \hat{\mathcal{H}}]\ne 0$ and does not satisfy 
the conditions (i) and (ii).  
Thus, if the system does not have the cross hopping satisfying $[\hat{W}, \hat{\mathcal{H}}]\ne 0$, 
the partial spin currents are prohibited by the Bloch-like argument, irrespective of whether 
the condition (i) and (ii) are satisfied.

\bibliography{excitonic.bib,bloch.bib,Paper.bib}
%\bibliography{paper.bbl}

\end{document}